# Observation of photon antibunching with a single conventional detector

Shaojie Liu,[1] Xing Lin,[1,3,a)] Feng Liu,[2,3] Hairui Lei,[1] Wei Fang[4] and Chaoyuan Jin[2,3]

1. Center for Chemistry of High-Performance & Novel Materials, Department of Chemistry, Zhejiang University, Hangzhou 310027, China
2. Interdisciplinary Center for Quantum Information, Zhejiang University, Hangzhou 310027, China
3. College of Information Science and Electronic Engineering, Zhejiang University, Hangzhou 310027, China
4. Interdisciplinary Center for Quantum Information, State Key Laboratory of Modern Optical Instrumentation, College of Optical Science and Engineering, Zhejiang University, Hangzhou 310027, China
a) Email: lxing@zju.edu.cn

**Abstract**

The second-order photon correlation function $g^{(2)}(\tau)$ is of great importance in quantum optics. $g^{(2)}(\tau)$ is typically measured with the Hanbury Brown and Twiss interferometer which employs a pair of single-photon detectors and a dual-channel time acquisition module. Here we demonstrate a new method to measure and extract $g^{(2)}(\tau)$ with a standard single-photon avalanche photodiode (dead-time = 22 ns) and a single-channel time acquisition module. This is realized by shifting the informative coincidence counts near the zero-time delay to a time window which is not obliterated by the dead-time and after-pulse of detection system. The new scheme is verified by measuring the $g^{(2)}(\tau)$ from a single colloidal nanocrystal. Photon antibunching is unambiguously observed and agrees well with the result measured using the standard HBT setup. Our scheme simplifies the higher-order correlation technique and might be favored in cost-sensitive circumstances.

The second-order correlation function $g^{(2)}(\tau)$ is fundamentally important in the development of quantum optics. It was introduced to analyze the time-dependent intensity fluctuations in a light beam, given by:

$$g^{(2)}(\tau) = \frac{\langle n(t)n(t+\tau)\rangle}{\langle n(t)\rangle\langle n(t+\tau)\rangle}, \quad (1)$$

where $n(t)$ is the photon number detected per unit time and $\langle\cdots\rangle$ denotes the time average over a long integration time[1]. A beam of light with $g^{(2)}(0) < 1$, representing photon antibunching, constitutes a new class of light sources which is unachievable in classical wave theory. For photons emitted by an ideal individual two-level system (TLS), $g^{(2)}$ goes to 0 at the zero-time delay. By using a Hanbury Brown and Twiss (HBT) setup (Fig. 1(a)), photon antibunching has been observed in many systems that can be treated as a TLS, such as individual atoms[2], molecules[3,4], quantum dots[5-10] and defects in crystal[11-14]. The major equipment cost in standard HBT setup comes from the two detectors, especially for systems working in the infrared wavelength range. In principle, the two detectors can be replaced by only one ideal single-photon-detector with no dead-time. The commercially-available state-of-art single-photon detectors and time acquisition module have a typical dead-time and after-pulse at the range of 2-100 ns, which is the time period that the detection system is blind to incoming photons after one photon is registered. Recently, remarkable efforts have been devoted to developing detectors with ultra-short dead-time (< 2 ns) such as superconducting nanowire detectors[15] and gated Geiger mode InGaAs avalanche photodiodes (APDs)[16]. Both photon bunching and antibunching have been observed using those novel detectors.

In this letter, we report a simple and low-cost approach to acquire the second-order correlation function using only one standard single-photon avalanche photodiode (SPAD) with a dead-time of 22 ns and a single-channel time acquisition module. Compared with previous photon correlation measurements based on a single detector[15,16], our scheme relaxes the requirement for the dead-time of the detector, which is crucial for TLSs with characteristic correlation time in the range of sub-nano to nano-second, which has been typically found in the studies of self-assembled QDs[9,10].

As a common implementation of the HBT measurement, shown in Fig. 1(a), the input photon stream is divided into two beams and coupled into two single mode fibers (SMFs) which guide photons to two individual SPADs. Time intervals ($\tau$) between two photon detection events that successively triggers the start and stop channels ($t_{\text{start-stop}}$) are recorded by a time to amplitude convertor (TAC). For short-enough time delay and low-enough photon counting rates, the $g^{(2)}(\tau)$ can be obtained by normalizing the histogram of time intervals to that of a Poissonian source and assuming it feeds the same number of photons into two channels[17,18]. Experimentally, an electrical delay ($t_d$) can be inserted to the stop channel to shift the optical zero-time delay to the positive axis in the 'start-stop' coordinate, explicitly:

$$t_{\text{start-stop}} \equiv \tau + t_d. \quad (2)$$

A comb of short peaks shown in Fig. 1(b) is a typical simulated $g^{(2)}$ function from a non-ideal TLS under pulsed excitation with a $t_d$ of 1500 ns. Peaks marked with '1−' and '1+' come from coincidence counts contributed by photon detection events from two adjacent excitation cycles. The suppression of the peak '0' reveals the antibunching of photon emission induced by a single excitation pulse. As a common practice, the ratio of the integrated area of the center and side features describes the extent of antibunching:

$$g^{(2)}_{0,\text{std}} \equiv \frac{\text{area of peak 0}}{\text{area of peak 1+ or 1-}} = \frac{A_{0,\text{std}}}{A_{1\pm}}. \quad (3)$$

Figure 1(c) shows the $g^{(2)}$ under continuous-wave (c.w.) excitation with a $t_d$ of 600 ns. The pronounced dip—$g^{(2)}_{\text{std}}(\tau = 0)$—reveals the same quantum nature of light as $g^{(2)}_{0,\text{std}}$ acquired in pulsed case. The red dash line indicates a typical criterion of $g^{(2)}(0)$ drawn at 0.5 for single-photon emitter. The non-zero $g^{(2)}_{0,\text{std}}$ and $g^{(2)}_{\text{std}}(0)$ originate from various factors such as the cascade emission of a non-ideal-TLS, the limited time resolution of electronics or the non-negligible laser background.

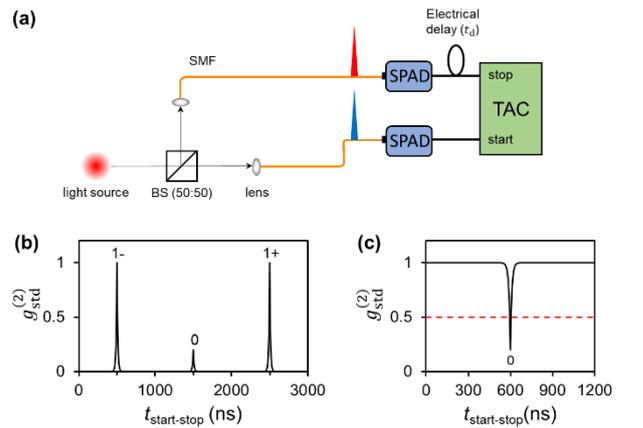

FIG. 1. (a) Schematic diagram of a standard HBT setup. A TAC working in the start-stop mode is used to analysis the time delay ($t_{\text{start-stop}}$) between adjacent detection events occurring in the start and stop channel. An electrical delay of $t_d$ in stop channel shifts the optical zero-time delay to the positive half of the coordinate. Simulated $g^{(2)}$ curves of a non-ideal TLS with lifetime of 10 ns under non-resonant 1MHz-pulsed (b) and c.w. (c) excitation with a $t_d$ of 1500 and 600 ns respectively show clear antibunching effect.

The reason to use two detectors in standard HBT setup is to circumvent the fairly long dead-time of detection system. Within a period of dead-time of detection, the histogram of time intervals in single channel must be zero, giving a fake appearance of antibunching. Furthermore, the after-pulse effect caused by charge carriers in the detectors released during the previous photon detection process disturbs the statistics of real photon detection time intervals.

Measuring the $g^{(2)}(\tau)$ with a single detector and a single-channel time acquisition module on the one hand reduces the cost of the setup, especially for the infrared band; on the other hand simplifies the setup. Previous measurement scheme based on a single detector employs detectors with ultra-short dead-time such as superconducting nanowire detectors or gated Geiger mode APDs[15, 16]. However, it is still not possible to get the $g^{(2)}(0)$ value directly in these cases due to the non-zero dead-time. Note that the core problem here is how to distinguish photon pairs emitted by the source with time intervals shorter than the dead-time of the detection channel. To solve this problem, we propose here a new scheme to separate the photon pairs by introducing an optical delay and develop a method to extract the second-order correlation function. Figure 2(a) shows the schematic of our setup to measure $g^{(2)}$ function using a single detector. We use a 50% : 50% beam-splitter (BS) to split the incoming beam just like the standard HBT setup and introduce an optical delay $\Delta t$ to 'store' one photon along one path. $\Delta t$ is chosen in the way that the photon in the longer path is dragged to bypass the abnormal period of detectors and timing electronics. Before being detected by the detector (Perkin Elmer SPCM-AQRH-16), two beams are made colinear through a knife-edge right-angle mirror. The absolute arrival time of photons in the merged beam is recorded by a time to digital convertor working in absolute-time mode (Becker & Hickl DPC-230). We can acquire a set of time intervals between adjacent photon registration events ($\tau \equiv t_{\text{adj.}}$) from absolute arrival times, base on which histogram of $t_{\text{adj.}}$ is calculated. Based on statistical analysis, second-order correlation function $g^{(2)}_{\Delta t}(\tau)$ in our $\Delta t$-delay scheme is related to the $g^{(2)}_{\text{std}}(\tau)$ measured in standard setup by Eq. 4:

$$g^{(2)}_{\Delta t}(\tau) = (R^2 + T^2)g^{(2)}_{\text{std}}(\tau) + RT g^{(2)}_{\text{std}}(\tau - \Delta t) + RT g^{(2)}_{\text{std}}(\tau + \Delta t), \quad (4)$$

where $T$ and $R$ is the overall counting ratio between path 1 and 2 (with $T + R = 1$). If $T = R = 0.5$, we have Eq. 5:

$$g^{(2)}_{\Delta t}(\tau) = \frac{g^{(2)}_{\text{std}}(\tau)}{2} + \frac{g^{(2)}_{\text{std}}(\tau - \Delta t)}{4} + \frac{g^{(2)}_{\text{std}}(\tau + \Delta t)}{4}. \quad (5)$$

Under periodic pulsed excitation, photons are emitted within a definite time window with uncertainty defined by the excited state lifetime ($t_{\text{lf}}$) of TLSs. In our scheme, photon pair emission events triggered by a single excitation pulse has a probability of 50% to generate histogram counts at $\Delta t$-delay (peak '0' in Fig. 2(b)). As long as $\Delta t$ is larger than the void time of detection channel and $t_{\text{lf}}$, photon pair information is preserved and shifted to the otherwise empty time-window.

Figure 2(b) shows the simulated $g^{(2)}_{\Delta t}$ under 1 MHz pulsed excitation with $\Delta t$ of 300 ns. Just like peak '1-' and '1+' in Fig. 1(b), peak '1L' and '1R' come from photon emission events of adjacent excitation cycle with one of whose detection time delayed by $\Delta t$ (each case of which occurs at a probability of 25%). Peak '1' in Fig. 2(b) is contributed by photons pairs going through the same optical path which is omitted in the standard two-detector start-stop scheme. Areas under the peak-triplet ('1L', '1' and '1R') have a ratio of $A_{1L}:A_1:A_{1R} = RT:R^2 + T^2:RT = 1:2:1$ and the periodicity decided by excitation laser is also preserved shown by the red shading for an example. Peak '0', '1L' and '1R' is equivalent to peak '0', '1-' and '1+' in the standard scheme. We can define $g^{(2)}_{0,\Delta t}$ using the ratio between peak '0' ($A_{0,\Delta t}$) and peak '1L' ($A_{1L}$) or '1R' ($A_{1R}$) and get the relationship between $g^{(2)}_{0,\Delta t}$ and $g^{(2)}_{0,\text{std}}$ using Eq. 6:

$$g^{(2)}_{0,\Delta t} \equiv \frac{A_{0,\Delta t}}{A_{1L/R}} = \frac{A_{0,\text{std}}}{A_{1\pm}} = g^{(2)}_{0,\text{std}}. \quad (6)$$

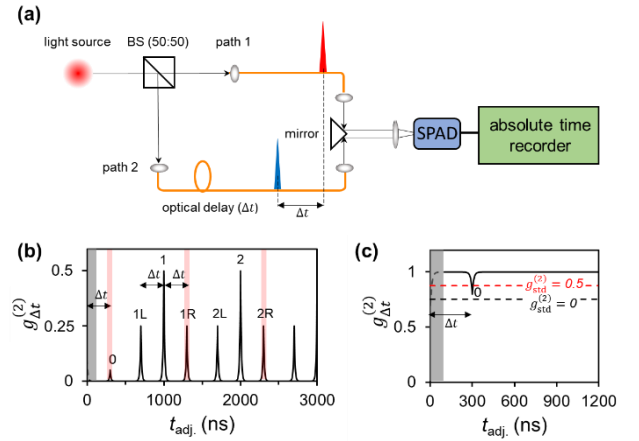

FIG. 2. (a) Experimental setup to measure second-order correaltion using one detector. An optical delay $\Delta t$ (300 ns, longer than the dead-time of detectors) is introduced along one of light paths. Time intervals of adjacent photon events ($t_{\text{adj.}}$) are obtained from the absolute time of photon registration events. (b) and (c) Simulated $g^{(2)}$ of a non-ideal TLS (10 ns lifetime) under 1 MHz-pulsed and c.w. excitation, taking the dead-time effect into account (light-gray shaded area) or not (black dash curve). At a time-delay of $\Delta t$, new features can be observed in both cases. Though coincidence counts around zero time-delay are obliterated by the dead-time effect, the replica at $\Delta t$ preserve a part of undisturbed coincidence events.

Figure 2(c) shows the simulated $g^{(2)}_{\Delta t}$ results under c.w. excitation with a $\Delta t$ of 300 ns. The case is more complicated because the coincidence counts at $\Delta t$ are not only the reflection of photon pair emission (one of which delayed by $\Delta t$) within the same excitation cycle but also contributed by single photon emission generated in two individual excitation cycles. Evaluating at $\tau = \Delta t$, Eq. 5 degenerates to give Eq. 7 with assumption $\Delta t \gg t_{lf}$:

$$g^{(2)}_{\Delta t}(\Delta t) = \frac{g^{(2)}_{std}(\Delta t)}{2} + \frac{g^{(2)}_{std}(0)}{4} + \frac{g^{(2)}_{std}(2\Delta t)}{4}$$
$$= 0.5 + 0.25\, g^{(2)}_{std}(0) + 0.25. \quad (7)$$

An ideal TLS with $g^{(2)}_{std}(0) = 0$ would give a $g^{(2)}_{\Delta t}(\Delta t)$ of 0.75 and the $g^{(2)}_{std}(0) = 0.5$ criteria in the standard HBT measurement has to be regulated to $g^{(2)}_{\Delta t}(\Delta t) = 0.875$ (Eq. 7), as indicated by the black and red dashed line respectively in Fig. 2(c).

When detection channel dead-time is taken into account, shown as the gray shaded area in Fig. 2(b) and (c), $g^{(2)}_{\Delta t}$ drops to zero around zero-time delay and cannot be acquired correctly, highlighting the importance of optical delay in our scheme.

To verify our single-channel photon correlation measurement scheme (see Fig. 2(a)), we measured the $g^{(2)}$ function of a single colloidal nanocrystal. Semiconductor nanocrystals, also known as colloidal quantum dots (CQDs), can emit anti-bunched single-photons. Due to its high stability[19, 20], easy fabrication and manipulation, CQD has been used in many proof-of-concept work of single-photon sources[21-23]. The synthetization of the CQD sample has been discussed elsewhere[24]. The PL emission of our ensemble QDs peaks at 625.6 nm with a FWHM of 29.6 nm at room temperature (Fig. 3(a) inset). The transient PL dynamics reveal a single exponential decay channel with $t_{lf}$ of 29.5 ns (Fig. 3(a)). The size of QDs used for measurement is ~14 nm, as shown in Fig. 3(a) inset. Samples for single-QD measurement were prepared by spin-casting a dilute solution of CdSe/CdS dot in plate CQDs in a PMMA/toluene (3% w.t.) onto a clean glass coverslip with a thickness of 0.17 mm. An epifluorescence microscope with suitable spectral filters combination was employed to excite individual QDs and collect the emitted photons. The excitation source was a 404 nm pico-second pulsed laser with 1 MHz repetition rate and 50 ps pulse duration. The excitation fluence density was 10 uJ/cm$^2$.

The $g^{(2)}$ function of the single CQD was first measured using the standard HBT setup equipped with two fiber coupled SPADs (Fig. 2(b)). Raw coincidence data are shown on the right-y-axis. The suppression of peak '0' at 1500 ns indicates the significant antibunching of single-photon emission of the single-QD. The central peak is nonzero due to non-ideal processes existing in CQDs as single-photon emitter. Spectrally overlapped bi-exciton emission degrades single-photon purity from exciton-emission at room temperature. Recent work shows that $g^{(2)}_{0,std}$ actually reflects PL quantum yield (QY) ratio between bi-excitons and single excitons[25]. From results in Fig. 3(b), the $g^{(2)}_{0,std}$ is calculated as 0.052 after carrying out background signal correction, which also implies the QY ratio of bi- and single exciton of the CQD under test.

We then measured the $g^{(2)}$ function of the same CQD using our new scheme illustrated in Fig. 2(a). The optical delay is achieved by an SMF delay line of ~80 m, corresponding to $\Delta t = 373$ ns. The combined beam is detected using a free-space SPAD.

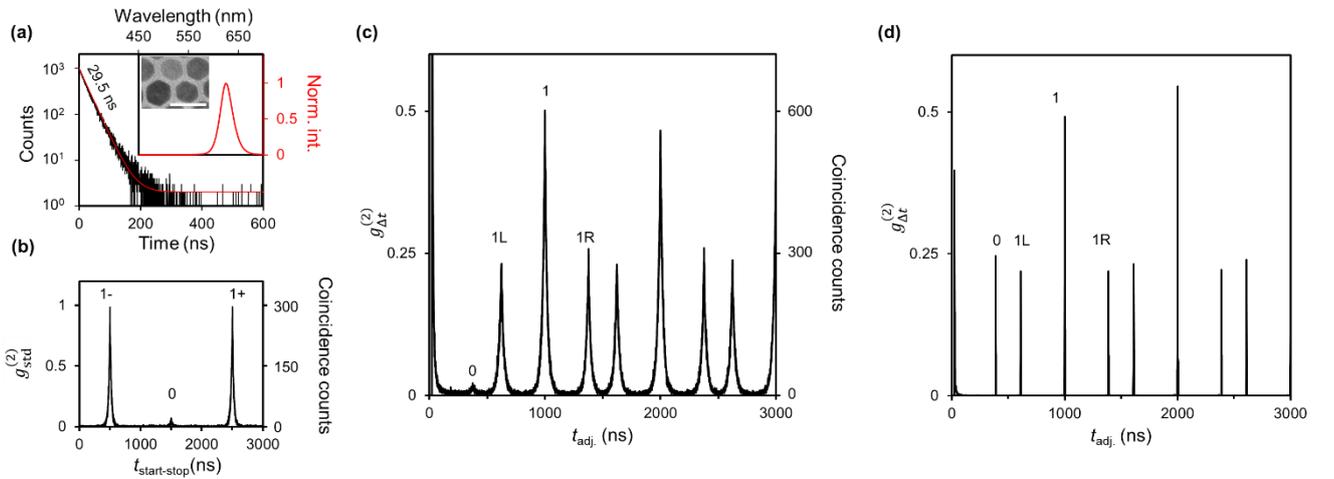

FIG. 3 (a) QDs under test show a single exponential decay with a lifetime of 29.5 ns. Black: experimental data; Red: fitting curve; Inset: photoluminescence spectrum and TEM, scale bar is 20 nm. Single-dot $g^{(2)}$ measurement under pulsed excitation for standard HBT (b) with 1500 ns electrical delay and one detector configuration (c) with optical delay of 373 ns. Data in (b) and (c) comes from the same single-QD with the same integration time. (d) $g^{(2)}$ measurement of pulsed excitation laser using single detector.

Optical path is well-aligned to ensure equal counting contribution from two paths and equal counting rate compared with standard scheme. The absolute time of photon detection events is recorded in real time and autocorrelation was performed off-line. The measured $g^{(2)}$ function is shown in Fig. 3(c). The value of $g^{(2)}$ vanishes in time delay range of 0-22 ns, which reflects the dead-time of our system. It is worth nothing that the value of $g^{(2)}$ after dead-time (22 -100 ns) is singularly high, which is attributed to the after-pulse effect of the detector. As mentioned above, the peak at $\tau = 373$ ns (labelled as '0') is the replica of central peak undisturbed by imperfection of detection channel. The observed strong suppression of '0' peak compared with peak '1L' or '1R' discloses the antibunching feature of the single-photon emitter. Quantitively, $g^{(2)}_{0,\Delta t}$ is 0.056 after background correction, which is in excellent agreement with the standard HBT measurement result (0.052). With the same integration time, raw coincidence counts under peak '1-/1+' in Fig. 3(b) and peak '1L/1R' in Fig. 3(c) are nearly the same, demonstrating the same data acquisition efficiency. To verify the validity of the antibunching signal, we use the same setup to measure the correlation of photons from pulsed excitation laser. As shown in Fig. 3(d), $A_{0,\Delta t}$ shares nearly the same value as $A_{1L}$ or $A_{1R}$, inheriting the randomness from coherent state, as expected.

We further compare the two methods under c.w. excitation (Fig. 4). A single CQD was excited by 404 nm c.w. laser with an intensity of ~130 W/cm$^2$. As for the standard HBT configuration, the second-order correlation shows a dip near delay of 600 ns, revealing the antibunching feature (Fig. 4(a)). After background correction[18], the $g^{(2)}$ at zero-delay time is 0.057. Figure 4(b) shows the results acquired using our new scheme with one detector. The value of shallow dip at $\tau = 373$ ns (dip '0') is 0.764. According to Eq. 7, $g^{(2)}_{std}(0)$ is determined as 0.056, in consistent with the one acquired from Fig. 4(a). $g^{(2)}$ measurement of the attenuated c.w. excitation laser using single-detector scheme shown in Fig. 4(c) is a flat line across all time delay except for the beginning 0-100 ns. The dead-time and after-pulse effect can be clearly observed in Fig. 4(c) inset.

We note that to obtain the $g^{(2)}$ function efficiently and correctly with our new scheme, the following factors should be taken into consideration: (a) Under pulsed excitation, photon counting ratio of $T:R$ between two light path plays an important role in determining the ratio of side replica to central peak, namely, how much coincidence events near zero-time delay can be shifted to undisturbed time window. The data acquisition efficiency, defined as ratio of shifted coincidence to the un-shifted one, reaches maximum of 50% when $T=R=0.5$. 50% efficiency is just the same as that in standard HBT setup (defined as the probability that photon pair been detected by two detection channels); (b) To avoid the overlap of shifted coincidence counts and un-shifted coincidence counts or one from different excitation cycle, a suitable optical delay should be several times smaller than the excitation pulse period, meanwhile being larger than the dead-time of detection channel and $t_{lf}$. Experimentally, this inversely sets an upper-limit to the excitation repetition rate; (c) Our configuration might be further simplified by utilizing a 2x2 fused fiber coupler to split the input beams and employing another 2x2 fiber coupler to combine two beams. All-fiber configuration becomes much more robust against variation in environment with the cost of losing photon counting rate due to fiber-connection and combination. Furthermore, the wavelength-dependent splitting ratio of fiber coupler should also be taken well care of.

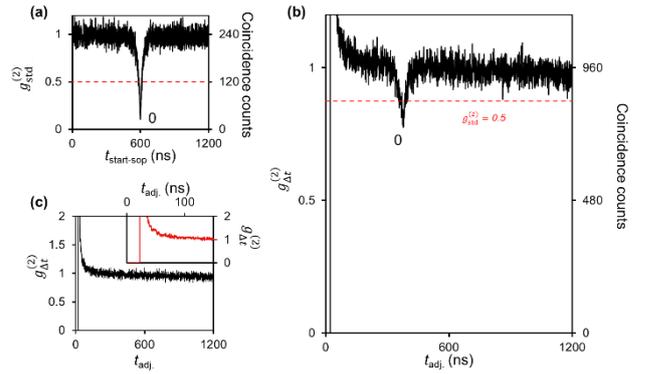

FIG. 4. (a) and (b) The $g^{(2)}$ measured under c.w. excitation. (c) The $g^{(2)}$ of excitation laser with single detector shows clear dead-time and after-pulse effect. Inset: Enlarged details in 0-150 ns.

In conclusion, we have demonstrated a simple and low-cost scheme based on a single conventional SPAD and single-channel time recording electronics to measure and extract the second-order photon correlation function. The new scheme is verified by measuring the $g^{(2)}$ function of a single photon emitter. Under both pulsed and c.w. excitation, our scheme reproduces the same information as the that obtained using the standard HBT setup. In principle, our approach might be applied to other high-order correlation measurement such as bunched light identification[26] and Hong-Ou-Mandel experiment[27].

**References:**

1. R. Loudon, *The quantum theory of light*. (OUP Oxford, 2000).
2. J. McKeever, A. Boca, A. D. Boozer, R. Miller, J. R. Buck, A. Kuzmich and H. J. Kimble, Science **303**, 1992-1994 (2004).
3. B. Lounis and W. E. Moerner, Nature **407**, 491-493 (2000).
4. J. Hwang, M. Pototschnig, R. Lettow, G. Zumofen, A. Renn, S. Gotzinger and V. Sandoghdar, Nature **460**, 76-80 (2009).
5. I. Sollner, S. Mahmoodian, S. L. Hansen, L. Midolo, A. Javadi, G. Kirsanske, T. Pregnolato, H. El-Ella, E. H. Lee, J. D. Song, S. Stobbe and P. Lodahl, Nat. Nanotechnol. **10**, 775-778 (2015).
6. N. Somaschi, V. Giesz, L. De Santis, J. C. Loredo, M. P. Almeida, G. Hornecker, S. L. Portalupi, T. Grange, C. Antón, C. Demory, C. Gómez, I. Sagnes, N. D. Lanzillotti-Kimura, A. Lemaître, A. Auffeves, A. G. White, L. Lanco and P. Senellart, Nat. Photonics **10**, 340-345 (2016).
7. H. Wang, Z. C. Duan, Y. H. Li, S. Chen, J. P. Li, Y. M. He, M. C. Chen, Y. He, X. Ding, C. Z. Peng, C. Schneider, M. Kamp, S. Hofling, C. Y. Lu and J. W. Pan, Phys. Rev. Lett. **116**, 213601 (2016).
8. M. Schwartz, E. Schmidt, U. Rengstl, F. Hornung, S. Hepp, S. L. Portalupi, K. Llin, M. Jetter, M. Siegel and P. Michler, Nano Lett. **18**, 6892-6897 (2018).
9. F. Liu, A. J. Brash, J. O'Hara, L. Martins, C. L. Phillips, R. J. Coles, B. Royall, E. Clarke, C. Bentham, N. Prtljaga, I. E. Itskevich, L. R. Wilson, M. S. Skolnick and A. M. Fox, Nat. Nanotechnol. **13**, 835-840 (2018).
10. J. Liu, R. Su, Y. Wei, B. Yao, S. Silva, Y. Yu, J. Iles-Smith, K. Srinivasan, A. Rastelli, J. Li and X. Wang, Nat. Nanotechnol. **14**, 586-593 (2019).
11. L. J. Rogers, K. D. Jahnke, T. Teraji, L. Marseglia, C. Muller, B. Naydenov, H. Schauffert, C. Kranz, J. Isoya, L. P. McGuinness and F. Jelezko, Nat. Commun. **5**, 4739 (2014).
12. P. C. Humphreys, N. Kalb, J. P. J. Morits, R. N. Schouten, R. F. L. Vermeulen, D. J. Twitchen, M. Markham and R. Hanson, Nature **558**, 268-273 (2018).
13. A. Sipahigil, R. E. Evans, D. D. Sukachev, M. J. Burek, J. Borregaard, M. K. Bhaskar, C. T. Nguyen, J. L. Pacheco, H. A. Atikian, C. Meuwly, R. M. Camacho, F. Jelezko, E. Bielejec, H. Park, M. Lončar and M. D. Lukin, Science **354**, 847 (2016).
14. Y. Zhou, A. Rasmita, K. Li, Q. Xiong, I. Aharonovich and W. B. Gao, Nat. Commun. **8**, 14451 (2017).
15. G. A. Steudle, S. Schietinger, D. Höckel, S. N. Dorenbos, I. E. Zadeh, V. Zwiller and O. Benson, Phys. Rev. A **86**, (2012).
16. A. R. Dixon, J. F. Dynes, Z. L. Yuan, A. W. Sharpe, A. J. Bennett and A. J. Shields, Appl. Phys. Lett. **94**, 231113 (2009).
17. M. Lippitz, F. Kulzer and M. Orrit, ChemPhysChem **6**, 770-789 (2005).
18. R. Brouri, A. Beveratos, J.-P. Poizat and P. Grangier, Opt. Lett. **25**, 1294-1296 (2000).
19. O. Chen, J. Zhao, V. P. Chauhan, J. Cui, C. Wong, D. K. Harris, H. Wei, H.-S. Han, D. Fukumura, R. K. Jain and M. G. Bawendi, Nat. Mater. **12**, 445-451 (2013).
20. H. Qin, Y. Niu, R. Meng, X. Lin, R. Lai, W. Fang and X. Peng, J. Am. Chem. Soc. **136**, 179-187 (2014).
21. B. Lounis, H. A. Bechtel, D. Gerion, P. Alivisatos and W. E. Moerner, Chem. Phys. Lett. **329**, 399-404 (2000).
22. P. Michler, A. Imamoğlu, M. D. Mason, P. J. Carson, G. F. Strouse and S. K. Buratto, Nature **406**, 968-970 (2000).
23. X. Lin, X. Dai, C. Pu, Y. Deng, Y. Niu, L. Tong, W. Fang, Y. Jin and X. Peng, Nat. Commun. **8**, 1132 (2017).
24. Y. Wang, C. Pu, H. Lei, H. Qin and X. Peng, J. Am. Chem. Soc. **141**, 17617-17628 (2019).
25. G. Nair, J. Zhao and M. G. Bawendi, Nano Lett. **11**, 1136-1140 (2011).
26. W. W. Chow, F. Jahnke and C. Gies, Light: Science & Applications **3**, e201-e201 (2014).
27. H. Kim, S. M. Lee, O. Kwon and H. S. Moon, Opt. Lett. **42**, 2443-2446 (2017).